\documentclass[10pt]{article}
\usepackage{courier}
\usepackage{graphicx}
\usepackage{hyperref}
\usepackage{xcolor}
\hypersetup{
	colorlinks,
	linkcolor={red!50!black},
	citecolor={blue!50!black},
	urlcolor={blue!80!black}
}
\usepackage{mathtools}
\usepackage[labelfont=bf,margin=1in,labelsep=endash,justification=raggedright]{caption}

\title{Tightly-Held and Ephemeral Psychometrics: Password and Passphrase Authentication Utilizing User-Supplied Constructs of Self}
\author{Christopher S. Pilson \\
	Braintrust Applied Research, LLC \\
	}
	
\begin{document}

\maketitle

\begin{abstract}
This research investigates the role of passwords and passphrases as valid authentication methodologies. Specifically, this research dispels earlier work that ignores information-theoretic lessons learned from cognitive and social psychology and psycholinguistics, and extends and enriches the current password security model.
\end{abstract}


\section{Introduction}
Computer security has traditionally been technology- or system-oriented. This approach has resulted in ingenious solutions to critical issues as user authentication, key distribution, and key expiration, but these solutions typically come yoked to new problems for computer users and administrative staff alike. In recent years, biometric measures have been gaining popularity for user authentication, but use of such devices may give rise to a \textit{false} sense of security: it has been possible to authenticate an artificial "finger" through use of various materials over the past few years. Materials and methods used range from "gummy" \cite{Matsumoto2002} or "putty" [Burt, 2004] products to mold a fingerprint, to use of cyanoacrylate ("CA", or super glue) \cite{ChaosComputerClub2004} and photo-lithography \cite{Matsumoto2002} to lift and reproduce latent fingerprints for later presentation. Unless exceptional care is taken with the planning and implementation process, "hard" biometrics as fingerprints present a significant security challenge. A better biometric has come in the form of "soft" biometric measures--typically user keystroke patterns. While this demonstrates a significant advance in methodological flexibility over "hard" biometrics, both user and system must undergo a training period in which the user becomes known to the system. This is a security system unlike a hard biometric system, as it offers easy key revocation; however, the key revocation and re-issue process here yields two similar keys, due to the typing characteristics of the end user. There must, then, be a system that is cognitively simple while retaining a reasonable security level and possessing revocability that will not compromise future keys.

Passwords and passphrases, when married with psychology and psycholinguistics, yield an authentication scheme that is revocable, memorable, and secure. This research demonstrates the potential effectiveness, simplicity, and security involved in this authentication system through an information-theoretic model that views authentication of a user as a \textit{shared secret} between user and machine. While this view is not new, the significant contribution that this work makes is the acknowledgement of the user's view of Self, and couples this with a password and passphrase selection process that serves to enhance the shared secret metaphor -- in effect, the system becomes a confidant for the user with respect to the chosen passphrase, and this helps ensure that a user will not violate this dyadic relationship, as the user has become invested in the protection of the shared secret.

\section{Methodology}
Successfully authenticating a user against a system has traditionally proven a difficult yet fruitful research area. From inception, user authentication has helped safeguard systems from unauthorized use. Traditionally, the requirement for doing so was one of cost efficiency, as legacy machines were expensive to operate and processor time was correspondingly expensive. However, in more recent years, the goal of user authentication has moved increasingly away from protecting a centralized resource and more towards safeguarding smaller computer systems, including personal computers, laptops, personal data assistants (PDAs), and cellular phones. The rise in decentralized and near-ubiquitous computing, coupled with the corresponding rise in machine interconnectedness and bandwidth across the public Internet, has increased the attack surface that each user on the network presents geometrically. As it is no longer uncommon for one user to possess multiple computer and network accounts, users are starting to feel somewhat overwhelmed with password policies that broker access to these connections. In effect, from an information theoretic point of view, password-based security systems are starting to decompose as further cognitive demands are placed upon computer users.

The reality of user-side information overload, along with the realization that there now exists a many-to-one relationship between targets and typical users, paints a much larger target on the back of computer users. This is especially poignant given that many users have a "preferred" password that they register, and later present, to each and every machine and network they contact. In this fashion, a dedicated attacker no longer has to break, extort, or guess many different authentication schemes, but only has to uncover the weakest one.

Authentication schemes provide a mechanism for a user to identify their presence to an attached security system. The classical mechanisms that are most immediately familiar to users in everyday environments are what the user knows, what the user \textit{has}, and what the user \textit{is} \cite{Meissner1976}, \cite{Wood1977}. "Modern" research tends to point away from the lowly password in favor of more recent authentication approaches; this research will not only examine password- and passphrase-based authentication, but present a model through which their security may be increased while simultaneously reducing cognitive load incumbent upon the end user.

As this research methodology is presented, it is important to note that no single system--even the information-theoretic model of passphrase-based security proposed here--is able to guard against every potential security breach. For instance, the first problem in password and passphrase assignment comes in the form of \textit{key distribution}. Needham \& Schroeder \cite{Needham1978} take some of the first steps in computing to formalize and secure the steps involved in key distribution across a network. However, the session key here is available for theft and, thereby, for attack. For instance, without the Denning-Sacco \cite{Denning1981} protocol extension, two principles may be eavesdropped upon by a third party through a replay attack. However, the Needham-Schroeder approach still provided for the first step towards protocol and key-exchange refinement, and provided early groundwork for Denning-Sacco and others to build upon. However, even the most robust distribution scheme and memory-compatible password presents little challenge to a shoulder-surfing thief. Hence, a series of specialized passwords--One-Time Passwords--are introduced. Harris's OPA \cite{Harris2002} mechanism allows a user the option of a \textit{disposable} password.

\paragraph{Passwords as an Authentication Methodology}
Passwords work on the principle that the security system shares a \textit{secret} with an authorized user--and this shared secret comes in the form of a password. Passwords are easy to employ, are not dependent upon anything external to the shared-secret concept to work, and can potentially present a low cognitive barrier of entry for a prospective user. However, policy must appropriately drive password use, lest this form of authentication present an easy target for an outside attacker. Appropriate password use is a phrase used here as a catch-all that describes user education and also conformity of the password to an accepted rigorous standard, where the rigor of a password is traditionally measured by either its ability to withstand attack or its conformity to a local policy.
 
The ability of a password to withstand attack is quantifiable through Anderson's Formula \cite{Anderson1972} (\eqref{eq:Andersons-Formula} on page~\pageref{eq:Andersons-Formula}), which gives the relationship between password complexity (measured in length and potential alphabet) and time required to carry out a brute-force attack. \textit{P}, in Anderson's Formula, represents the probability of a successful attack. Typically, Anderson's Formula is utilized wherein \textit{P} is set to 0.50, as this value serves as the mean run-time, $\Theta$, between $\Omega$ and O \cite{Knuth1997}.

\begin{equation} \label{eq:Andersons-Formula}
\text{P} \geq \frac{(\text{number of seconds})\times(\text{number of guesses per second})}{(\text{number of characters in password alphabet})^{\text{password length}}}
\end{equation}

Using this method, it becomes entirely possible to algebraically run this analysis in reverse to examine how long--and, therefore, how "complex"--a password ought to be to measure the average length of time it would take an attacker to break the password using brute-force guessing techniques.
 
With respect to standards- and policy-compliance issues, a 1985 publication by the United States Department of Defense \cite{U.S.DepartmentofDefense1985} serves as the "gold standard" of password complexity available to a user. In this document, the suggestion that user passwords be comprised of no less than 6 characters is made. Additionally of note--and often overlooked in practice--is the allowance for "user friendliness" in the form of pronounceable passwords, as these passwords are often easier to remember than entirely unpronounceable passwords. Indeed, such an approach is entirely possible and conceptually simple through use of phonemes rather than a more traditional alphabet.

\begin{figure}[h!]
 \caption{Simple Relationship Between Password Complexity and Overall Security}
 \centering
 \includegraphics[scale=1]{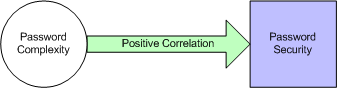}
 \label{fig:Complexity-to-Security}
\end{figure}

Using these traditional views, then, a secure password ought to be exceptionally long (which serves to defeat potential brute-force attacks through greatly increasing the search space and, therefore, the total run-time via a mathematical relationship between the two \cite{Anderson1972}). This is diagrammed in figure~\ref{fig:Complexity-to-Security} on page~\pageref{fig:Complexity-to-Security}. 

\begin{figure}[h!]
 \caption{Model of Overall Password Security, Post- Chunking and Recall}
 \centering
 \includegraphics[scale=1]{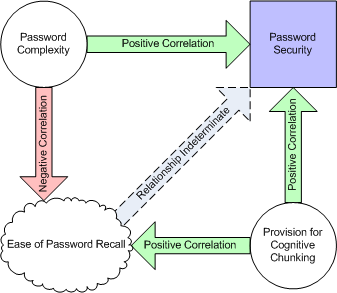}
 \label{fig:Chunking-and-Recall}
\end{figure}

However, increasing the password length with no regard for achieving parsimony with what is known about chunking \cite{Miller1956} from cognitive psychology inappropriately adds to the overall complexity of the password and, consequently, makes the password more difficult for an end user to remember. This lapse in memory, in turn, may lead to the end user writing down the password, thus overcoming any additional security performance gain picked up through use of a more complex password scheme. Therefore, it becomes apparent that the relationship between provisioning for cognitive chunking and overall password security is positive. The temptation, however, to comply with traditional "password wisdom" and suggest that ease of recall and password security are at odds with each other is refuted; instead, ease of recall is demonstrated as a proxy measure. This emerging model is demonstrated in figure~\ref{fig:Chunking-and-Recall} on page~\pageref{fig:Chunking-and-Recall}.

\paragraph{Pass-phrases: Beyond Password Authentication} 
Capitalizing on the earlier systems' success through the addition of some meaningful "content-rich" token to accompany the password itself to be used as a mnemonic device of sorts (e.g. the Department of Defense \cite{U.S.DepartmentofDefense1985} guideline suggesting that users verbalize the password into something more immediately accessible, chunkable, and memorable), Yan et al. \cite{Yan2004} suggested the creation of strong passphrase-like strings. These strings would be built out of cognitively simple and meaningful content, such that even a "gibberish" password of "MDFNIR" can be created out of the initial sentence "My dog's first name is Rex."

This is an important point to bring to bear in a password system, as these kinds of passwords are not found in a dictionary, and also have a high degree of entropy. And these are also precisely the conditions that make for fast pro-active password scanning \cite{Yan2001} to actively improve the quality of accepted passwords. In research, it has long been 
postulated that a \textit{flexible} approach be taken with respect to user passwords--however, as is typified by Saltzer \cite{Saltzer1974}, in the same breath that flexibility is promoted, so too comes the view that this notion itself is negatively correlated with security. For instance, in Saltzer's research, \textit{total mediation} is given as a necessary condition for security, which does away with the notion of credential caching, thus meaning that a user would have to type in a password many times in one computing session to gain full access over their virtual domain.

For all the ease and simplicity that passwords give users and security analysts, the fact remains that many users are given minimal instruction on password selection. As such, individuals tend to pick passwords that fall into various "classes"--an assertion uncovered by Petrie \cite{Petrie2001} and later re-tested "in the wild" on an e-commerce site \cite{Medlin2005}. All passwords used on this system by end users fell into one of 7 distinct categories, or a catch-all "other" category. 

We have seen, above, how the use of a \textit{passphrase} can act as an add-in to the more traditional \textit{password}, and this kind of marriage can result in passwords with a high degree of entropy. Moreover, as passphrases may better resist being forgotten \cite{Keith2005}, it seems that these "extended-length passwords" make a good trade-off between security and psychological acceptance for users and system architects alike. 

It is this aspect of passwords and passphrases, \textit{resistivity to forgetting}, that is of particular interest. To keep password load at a minimum, for example, Kurzban's cognitively simple and parsimonious system \cite{Kurzban1985} may be utilized. If, for instance, the fictitious user introduced at the start of this paper had to memorize and use several different passwords or passphrases for different authentication sessions, this task would be made vastly easier through the structure given here. 

\paragraph{Biometrics--A More Secure Alternative?}
In discussing user authentication, it is tempting to suggest \textit{more secure} alternatives to password- and passphrase-based authentication, as passwords and passphrases are not novel, and technology has clearly evolved since their inception. Biometrics have recently come into vogue as an "obvious" solution to the problem of user authentication as they are proposed as being cognitively transparent while simultaneously leveraging the "what the user \textit{is}" aspect of authentication. However, biometrics have their own dark secrets to harbor from the security community. 

Biometrics, it should not be forgotten, still suffer from key distribution issues that plague traditional security approaches--this fact is endemic in that no amount of protocol design will do away with the fact that the two agents involved--computer and user--must somehow be assured that the other exists and that the other is indeed who they claim. The typical exemplar of biometric systems, widely available for consumer and business use, is fingerprint recognition. In 2004, Uludag \& Jain \cite{Uludag2004} presented work in which fingerprint biometric systems were utterly broken and accepted an invalid principal as valid. Thus, breaking biometric systems is within the realm of reason, and in fact these systems promote other, more insidious problems while hiding behind an artificial cloak of security.

Bruce Schneier \cite{Schneier1999}, in his comparison of traditional and biometric security devices, elucidates and pierces this cloak. Using biometric measures such as fingerprints, he argues, suffer from lack of secrecy and key re-issue problems. Lack of secrecy in fingerprint scanners is a straight-forward concept, as the case might arise in which a user's fingerprint becomes compromised due to the fact that the oils leave a "residual" pattern upon the scanning glass of the biometric control device. This residual pattern, in fact, may be picked up with relative ease through use of techniques ranging from the very simple employ of facial mask gels and gummy bears to more advanced techniques involving photo-lithography. To this end, the author has aided in and examined the former method employed successfully by a non-expert. This problem, however, pales in comparison to the one of key re-issue. If a password or passphrase becomes compromised, it is reasonably simple to re-issue the user a new key. However, in the event that a fingerprint is compromised, it is impossible to "re-issue" the user with another dominant-hand index finger; the alternative is to use another digit, giving 10 possible "re-issues" of keys. In this vein, \textit{re-identification} also becomes problematic in system design, as again the number of keys is physically constrained at 10 per user. 

Users, Mike Just argues \cite{Just2004}, ought to 
\begin{quote}
ha[ve] the flexibility and choice to control the personal information, if any, that they provide [in their application for, or use of, an authentication token].
\end{quote}

Use of typical biometrics, especially fingerprint biometrics, takes away this control of personal information by instead physically depositing this information upon the fingerprint reader itself. Hence, through use of a \textit{direct} key (e.g. biometric token) rather than a \textit{proxy key} to be used to identify the user (e.g. passwords, passphrases, or another secret), the user loses flexibility and control in use of personal information. Even though biometrics have been touted--validly--as being more transparent to the end-user, there is no such thing as a "free lunch" in security. Rather, within any authentication system lies certain trade-offs and assumptions, and this point is examined best by O'Gorman \cite{OGorman2003}, who presents a simple and clean analysis of different security systems.

"Soft" biometrics, however, are not excluded in the same manner as the above static, so-called "hard" biometric measures that depend upon the user \textit{being} something fixed and unchanging. Rather, "soft" biometrics make use of a proxy measure, such as typing patterns, to identify a user. This field is still emerging, but holds great promise as it can implicitly provide for a "two-phase" approach wherein a user still \textit{actively} authenticates with a password or passphrase, and \textit{passively} authenticates using typing patterns. A solid analysis comes in Figure 1 of Peacock et al.'s \cite{Peacock2004} research, which performs a mini meta-analysis (by graphing false acceptance/rejection rates from other authors working in-field) on the data presented about keystroke biometrics, and lays it out for the reader in a reasonable and simple manner. As such, it becomes possible to zero in on approaches that perform well, and (re-)examine approaches that are non-confirmatory to the other in-group studies. Likewise, Figure 2 in this same research contains the cost \textit{to a user} to "enroll" in the systems presented in Figure 1. Keystroke biometric measures tend to make use of engines that allow flexibility within the system, and these approaches have included the use of Markov Models \cite{Chen2004} and other data-mining and machine-learning approaches, and the rhythm and cadence of typing \cite{Guven2003}, \cite{Bergadano2002}. 

\paragraph{Beyond the Fingerprint and Keyboard--Other "Soft" Biometrics}
In attempting to escape passwords and passphrases, recent security research has posited that perhaps \textit{image presentation} would achieve the goal of user authentication. In this type of authentication, a user would pick out an image from a collage and identify it as "their" image. However, unless image presentation is properly randomized, the initial selection process of an image may well be doomed, as research indicates handedness as a primary indicator of directional choice in a T-maze \cite{Scharine2002}. While this may appear to have nothing to do with security, this work extends and enriches the work by Robinson \cite{Robinson1933} that dealt with museum patrons. Biological \cite{Lund1930} explanations are not the cause of this preference, nor are visual cues (\cite{Blumenthal1928}; \cite{Brigden1935}; \cite{Howard1966}; \cite{Szymanski1913}) in use. 

However, Abed \cite{Abed1991} measured the visual scanning (e.g. reading) patterns of readers from different cultures, each having differing text directionality. Western participants demonstrated more left-right saccades than members of cultures wherein language is written top-to-bottom or right-to-left. In effect, this implies that there exists a \textit{non-random} component to saccades when individuals examine a computer monitor. This fact may be leveraged by an attacker examining a visual authentication system that does not make use of randomized image locations. Finally, it is the case that visual target acquisition presents a unique problem in human factors research, and that "sundries" such as monitor type (e.g. Cathode Ray Tube (CRT) versus Liquid Crystal Display (LCD) technology) may skew the visual search performance of a participant \cite{Hollands2002}, which will then introduce a bias into the image selection process.

\paragraph{Lessons Learned from Cognitive Psychology}
Thus far, passwords and passphrases have been examined only from a viewpoint of maximizing information retrieval based upon psycholinguistic cues. Just \cite{Just2004} flags \textit{applicability} and \textit{memorability} as important measures of usability in a challenge-response authentication framework, and this research has covered the latter thus far. Yet, there exists another "hook" into enhancing memory and recall of a target, and this forms the crux of the model's development--this notion, found in cognitive and social psychology, is known as \textit{self-reference effect}.

In \textit{self-reference and the encoding of personal information} \cite{Rogers1977}, Rogers et al. tested subjects' memory using lists of adjectives. Recall was greatly improved when, as demonstrated in Heatherton et. al. \cite{Heatherton2004}, subjects were given the word \textit{happy} and this prompt was referenced back to the participant--\textit{does} [the word] \textit{happy describe you?}--as opposed to being prompted \textit{does} [the word] \textit{happy mean the same as optimistic?} This self-referential type of encoding results in a more robust memory of the target--possibly because of the initial depth of processing while the user is trying to reconcile this target with the user's view of Self. Obviously, there would seem to exist a certain "latch" into creativity, imagination, the visuo-spatial sketchpad, and simple recall when trying to determine how like a target the end user is, or (more appropriately), how this target integrates with the end user and within their life experience. This study forms the basic premise of the expanded password security model presented here: that there is something inherently special about self-referenced prompts that vastly improves their recall through their tie back to the user.

Following this initial probe, Kuiper \& Rogers \cite{Kuiper1979} found that memory recall is enhanced for trait adjectives that are self-referent in nature, \textit{even when judgments about them are not made at the time of presentation}. This is important, as it added to the earlier work through presenting this as more than a "one-time-party-trick". Rather, the notion of the Self is so utterly important and pervasive that it would appear to be a full-blown and hard-wired system within the mind/brain that is always scanning, always looking for ways to integrate the world into itself, and itself into the world.

It is not enough, ideally, that a passphrase should only provide high recall for the user; rather, the passphrase itself ought to present a barrier to being revealed to a third party by the user. In many organizations today, users are left with a devalued sense of password and passphrase importance. The realization that this one "key" can be presented to the system by a foreign user who may then utilize the target's information for gain tends to be less immediate than the knowledge that conforming to password and security policy is largely painful for end users. These end users have no real "connection" to their passwords, thus users have repeatedly proven that these passwords are, in practice, worthless. Indeed, in a survey conducted at a security conference (USENIX), 70\% of respondents happily gave their passwords away for a bar of chocolate, a pen, or other cheap baubles of remuneration. An unscrupulous attacker could then leverage their newly-gained system access and take, alter, or destroy both data and infrastructure. Clearly, this kind of dissociation from passwords and passphrases (or any other kind of "token" presented to a system for authentication) creates very real and critical security issues for an organization.

\textit{Secrecy Helps People Maintain Their Personal} [Electronic] \textit{Boundaries}. This section heading in Kelly \& McKillop's \cite{Kelly1996} research on revealing personal secrets, when taken in context with the task at hand, provides a platitude that will light the stage and provide the ambiance for the real meat of this research, which is precisely \textit{how} to go about maintaining and preserving personal boundaries through electronic secret-holding. It is simple (and common) enough to assert that "a system is most secret when all its members act in secretive manners", but this assertion provides little in the way of actual and substantive value. In this research, I have been working on a common theme--the dyadic relationship between human and computer. This relationship inherits elements from human-human relationships, including \textit{trust} [both in the human and computer security implications of the word], [the potential for both sides to exercise] \textit{poor judgment}, and \textit{discretion}. All these elements, of course, surround a common theme; the human actor and the computer share a common secret, and are in this way confederates. Moreover, it becomes a not unnatural connection to discover that, just as individuals are psychologically geared to feel "relief" and "a load taken off their shoulders" upon divulgence of a secret, so too does this notion follow through to user-attacker human dyads. If, as Kelly \& McKillop point out, it is the case that disclosure and personally damning nature of the secret are directly related, then it stands to reason that a "strong" secret would be one that would utterly compromise or otherwise violate or intrude upon the individual involved. Hence, the original assertion that I made, in which I stated the seemingly intuitively obvious point that users would likely feel a stronger "tie" with passwords and phrases that were tightly coupled within their realm of experience than random passwords and phrases, has support.

This support at first appears only partial, yet when the \textit{computer} is brought on as the compliment to the individual, then the picture falls into place very rapidly. It is clear, if users can so completely allow themselves to \textit{interact} with a machine - without accepting anthropomorphic fallacy as the root cause--as is demonstrated in Nass \& Moon's work \cite{Nass2000}--then so too is it entirely within the realm of logic and reason that the users are carrying out a social interaction with the machines they contact. What this rejection of anthropomorphic fallacy obviates is an explanation into \textit{why computer users think their monitor, mouse, keyboard, and computer are ascribing human attributes to these bits of plastic, glass, and metal}. This is a good direction to take, as it is clear that computer users do not "think" of their computers as human or human-esque, but it is still the case that there exists a \textit{human-like interaction} between user and system. In effect, through using passwords and passphrases--as with any authentication scheme reliant upon a shared secret--users implicitly trust the machine and associated paraphernalia to \textit{maintain the shared secret, no matter how many times an attacker asks the machine to divulge it}. The security community, however, appears to be doing a great injustice to users and security architects alike through not first \textit{finding}, then \textit{explaining}, and finally \textit{encouraging} this form of interaction between security principal (user) and device (machine). 

\section{Finalizing the Proposed Model}
To this point, the research has focused primarily upon providing support for passwords and passphrases as a valid avenue of user authentication, even in light of newer technologies. Through the leveraging of cognitive psychology, it is possible to bind a token that is both \textit{personal and important} while still providing only a \textit{proxy} for access to a specific user. In this way, system access can remain secure in the aspect that access is brokered through a \textit{shared secret} between user and machine, while still upholding tenets of design presented in Mike Just's research. It is important, too, to make note of the fact that designing security systems should not be a static process; rather, as Yee \cite{Yee2004} points out, the security system may instead be designed in a manner similar to the approach of the Software Development Lifecycle (SDLC). A proponent of SDLC, Enger \cite{Enger1981} presents five distinct and discrete stages of development that a system should traverse before release. These stages provide a structured framework for systems development within which revisions, reviews, and feedback may take place. However, this framework acts as rigid scaffolding surrounding the work, and is therefore not necessarily the best approach to take or model in all instances. In Yee's research, a sort of "amoebic" cognitive model is employed that demonstrates the acceptable action points (e.g. states in a Finite State Machine) that are available to a user of the system. This "amoebic" approach is actually quite cognitively correct, as security has been talked about as "engulfing" or "encompassing" different target points and/or users. \textit{Security by admonition} and \textit{security by designation} are brought out as topics of system design. What makes this approach compelling--and potentially lucrative--is that this parallels the research into Leader-Member Exchange (LMX) modeling (the evolution of "dyadic relationships" in management) that was initially founded by Bass \cite{Bass1985}, and later extended \cite{Bass1998}. Hence, it is preferable to enforce security through designed control than through admonition and rigid, over-reaching, or potentially draconian policies. This approach will receive the same kind of "employee buy-in" that its analogue in LMX--Transformational Leadership \cite{Bass1998}, \cite{Bass2003}--receives, and this notion of control is preferable to policy-driven approaches of system security via passwords or passphrases in that it effectively brings the \textit{system} in as the second acting principal in what effectively becomes a dyadic relationship between user and machine. This is an effective strategy for preventing user password or passphrase divulgence, as this security-as-relationship view brings to bear secret-keeping with respect to oneself. Again, as a model of Self becomes increasingly more integral in the key presented to a principal, the ramifications of disclosure become magnified, such that a "security" breach--a user giving out their password or passphrase to an un-trusted party outside the system-user dyad--transmutes instead into divulgence of a personal secret.

A dyadic relationship, taken in a human-human construct, leaves itself open to plausible deniability thwarting the fear involved in a particularly outlandish or unlikely secret being revealed. However, it is important to take note of the fact that \textit{with a computer system acting as the second principal in a dyadic relationship}, plausible deniability is removed as a potential confound from the system, as well as the overall nature of secret-telling in life: ephemerality. For instance, all that one has to do to dispute that a human-borne secret "belongs" to them is either to simply deny that 
the other party has knowledge, or better still, to deny the secret itself. It is this notion of ephemerality that gives human-human trust dyads the confound of later repudiation: the value of holding a secret is largely nullified if its veracity cannot be determined by an outsider. Hence, the value of \textit{keeping} that secret is also decreased, as its potential for damage is mitigated by its transparency, especially in a dyad possessing an inequity between parties (e.g. employer-employee relationships).

\section{Discussion}
\begin{figure}[h!]
 \caption{Full Model for Password Security}
 \centering
 \includegraphics[width=\linewidth]{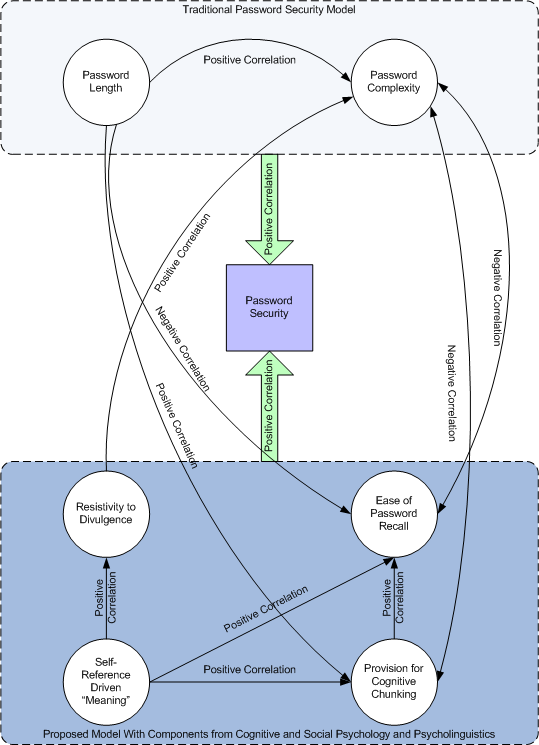}
 \label{fig:Full-Model}
\end{figure}

In this research, I have provided links between overall password/passphrase security and different factors, such as password length, password complexity, ability of a user to recall their password, meaningfulness of the password, and resistivity to secret divulgence through use of highly personal and "sensitive" information. Tying this information about the user's Self to the underlying construct of non-repudiation gives the finalized model displayed in figure~\ref{fig:Full-Model} on page~\pageref{fig:Full-Model} (with repudiation, as is seen in human-human secret-keeping dyads, a "repudiation" node would be present and would be negatively correlated to the security of the resulting secret--here, a password or passphrase).

We understand from literature that there exist consequences for secret disclosure, yet also that individuals are prone to tell secrets as a cathartic experience. However, when the secrets involve central constructs and components of the Self, and with removal of both repudiation and the "prisoner's dilemma" (as is the case when the secret confidant is a computer system rather than another human), secrets are both better protected and more potentially disastrous, should they be released. What this means for computer security is that the answer to user sharing and divulgence of shared secrets--thus providing access into the system itself to an unauthorized party--comes not at the price of policy and control, but more distributed and self-governing methodologies that tie into constructs of Self.

Coupled with this approach is the view that a user and computer system share a \textit{dyadic and symbiotic relationship}; system and user share a secret and hold this secret in trust. As this secret is representative of a tightly-held and ephemeral \textit{psychometric} rather than an observable biometric, this approach is secure from attack by inspection or release.

\section{Conclusion}
While there has been a recent surge of interest away from cumbersome and complex authentication schemes, research focus has tended away from 
such "simple" mechanisms as passwords and passphrases. This tendency is perhaps premature and in error; at AMCIS 2005, two interesting presentations were given that addressed both the \textit{selection} \cite{Medlin2005} and \textit{effectiveness} \cite{Keith2005} of passwords and passphrases. Password and passphrase selection has been described as falling into \textit{functional categories} (as in \cite{Medlin2005}, extending \cite{Petrie2001}). However, to date, the effect of social and cognitive psychology on password selection has been lacking, and psycholinguistic examination has been largely relegated to phoneme selection. This research demonstrates the potential for future and refined study of passwords and passphrases as a secure shared secret authentication mechanism when taken with the methodological framework given.

With the adoption of this framework, employee disclosure of system-access keys would diminish, as would employee requests for password and passphrase re-issue. Contrast this environment against the more traditional wrapper for securing passwords by making them unintelligible through policy. This results in vast resources being required to maintain such a system--after all, it is difficult to remember a 30-character mixed-case, alpha-numeric string containing special characters as \texttt{Xf033\_99!YU\&XmqZPl\^{}3j2Yxv4FPw7}. Given this, employees either resort to defeating system security completely through use of handwritten and posted notes at their workstation, or make use of corporate "help-desk" functions to re-issue their password--which, in turn, is subsequently either forgotten or written down. Add to this the current slew of policies that mandate password expiration after one to six months, and this issue becomes magnified further. The best scenario painted here is one in which an employee \textit{only} writes their password down, or \textit{only} calls a dedicated help-desk for assistance--the \textit{worst} scenario would come with employees sharing password and login information (which is then likely written down); if \textit{everyone} has the login information, then the nuisance of calling a help-desk is avoided. However, this completely and utterly breaks the system with respect to its ultimate goal of matching \textit{access} and \textit{user}, and being able to constrain and bind the two facets of use to a common key (here, a password or passphrase) through appropriate policy. 

This paints a bleak picture; passwords and passphrases, however, can be secure, memorable, \textit{and} impervious to release. A 30-character strong password with high entropy was given above; the passphrase
\begin{quote}
The first time I went to the beach in Florida was great; with the sun, sand, and surf, I felt like Camus!
\end{quote}
not only exceeds entropy and character counts, but is exceptionally "chunkable" and cognitively lightweight, as it provides an event that latches into some notion of Self about the user. As demonstrated, this marriage is good for the environment in which the system is placed, good for the end user, and good for security.

\bibliographystyle{ieeetr}
\bibliography{Tightly-Held-and-Ephemeral-Psychometrics}

\end{document}